\documentclass[
prc,%
10pt,%
final,%
notitlepage,%
oneside,%
twocolumn,%
nobibnotes,%
nofootinbib,
superscriptaddress,%
floatfix,%
showkeys,%
showpacs]%
{revtex4}
\usepackage{color}
\usepackage{amsfonts}
\usepackage{amsbsy}
\usepackage{mathrsfs}
\usepackage{graphicx}
\def\lsim{\mathrel{\rlap{
\lower4pt\hbox{\hskip-3pt$\sim$}}
    \raise1pt\hbox{$<$}}}     
\def\gsim{\mathrel{\rlap{
\lower4pt\hbox{\hskip-3pt$\sim$}}
    \raise1pt\hbox{$>$}}}     

\begin{document}
\title{Elliptic Flow of Protons and Antiprotons in Au+Au Collisions\\
at $\sqrt{s_{NN}}$ = 7.7--62.4 GeV within Alternative Scenarios of Three-Fluid Dynamics
 } 
\author{Yu.B. Ivanov}\thanks{e-mail: Y.Ivanov@gsi.de}
\affiliation{Kurchatov Institute, 
Moscow RU-123182, Russia}
\begin{abstract}
Analysis of elliptic flow of protons and antiprotons in Au+Au collisions
is performed   in a wide range of incident energies  $\sqrt{s_{NN}}$ = 7.7--62.4 GeV. 
Simulations has been done within the three-fluid model
employing a purely hadronic equation of state (EoS) and two versions of the EoS
involving deconfinement transition: 
an EoS with the first-order phase transition  
and that with a smooth crossover transition.
It is found that the proton data are  reproduced approximately to the same extent
within all of the scenarios, including the hadronic one, while  
the deconfinement scenarios look certainly preferable for the antiproton elliptic flow. 
The fact that difference between elliptic flows of protons and antiprotons decreases with 
the incident energy rise is a consequence of reducing baryon stopping rather than an
onset of  deconfinement. 
\end{abstract}
\maketitle

\section{Introduction}

The large azimuthal anisotropic flow at
the Relativistic Heavy Ion Collider (RHIC) is believed to be 
a conclusive evidence for the creation of dense partonic matter
in ultra-relativistic nucleus-nucleus collisions. 
This anisotropy
is described by flow parameters defined as
the proper Fourier coefficients $v_n$ of the particle distributions 
in azimuthal angle with respect to the reaction plane \cite{Voloshin:1994mz}. 
The major bulk of  research both theoretical and experimental has been done 
on the elliptic flow ($v_2$).

Recently new data of the STAR Collaboration \cite{Adamczyk:2013gw}
on transverse-momentum dependence of the elliptic flow of 
identified particles in the incident energy range of 
$\sqrt{s_{NN}}$ = 7.7--62.4 GeV were reported. These data were taken 
within the  Beam Energy Scan (BES) program proposed at RHIC. 
The BES program was proposed in order to study 
possible onset of deconfinement, as
well as possibly to identify the location of the critical end
point that terminates the crossover transition at small
quark-chemical potential to a first-order phase transition
at higher quark-chemical potential \cite{Aggarwal:2010cw}.

This work aims to analyze the new STAR data \cite{Adamczyk:2013gw}
within different scenarios (with and without 
deconfinement) of heavy-ion collisions in order to draw conclusions 
on the matter produced in these collisions. 
The simulations were performed within a model of the three-fluid 
dynamics (3FD) \cite{3FD} employing three different equations of state (EoS): a purely hadronic EoS   
\cite{gasEOS} (hadr. EoS) and two versions of EoS involving the deconfinement 
transition \cite{Toneev06}. These two versions are an EoS with the first-order phase transition
(2-phase EoS) and that with a smooth crossover transition (crossover EoS). 
Details of these calculations are described in Ref. \cite{Ivanov:2013wha}. 
First results of the 3FD simulations within alternative scenarios of heavy-ion collisions
are reported in Refs. \cite{Ivanov:2013wha,Ivanov:2012bh,Ivanov:2013yqa}
dedicated to analysis of the baryon stopping and particle production.  
Results on transverse flow in the energy range of AGS 
(Alternating Gradient Synchrotron) and SPS (Super Proton Synchrotron) but 
only within hadronic EoS were presented in Refs. \cite{3FDflow,3FDv2}.

The 3FD model \cite{3FD} treats the process of a nuclear collision from 
the very beginning, i.e. from the stage of incident cold nuclei, to 
the final stage of freeze-out. 
Contrary to the conventional hydrodynamics, where local instantaneous
stopping of projectile and target matter is assumed, a specific
feature of the 3FD is a finite stopping
power resulting in a counter-streaming regime of leading
baryon-rich matter. The basic idea of a 3-fluid approximation to
heavy-ion collisions \cite{Ivanov:1987wv,I87,MRS91} is that at each space-time
point a generally nonequilibrium 
distribution of baryon-rich
matter can be represented as a sum of two
distinct contributions
initially associated with constituent nucleons of the projectile
(p) and target (t) nuclei. In addition, newly produced particles,
populating the mid-rapidity region, are associated with a fireball
(f) fluid.
Therefore, the 3-fluid approximation is a minimal way to 
simulate the finite stopping power at high incident energies.

The main observation of Ref. \cite{Adamczyk:2013gw} is 
a strong difference in
$v_2(p_t)$ between particles and their corresponding antiparticles. 
It is argued that it cannot be explained in a purely
hydrodynamic approach since particles and anti-particles
have the same mass.
The evolution of the elliptic flow in the transient BES energy range 
was addressed in a number of theoretical works
\cite{Song:2012cd,Xu:2012gf,Steinheimer:2012bn,Dunlop:2011cf,Greco:2012hh,Konchakovski:2012yg}.

In Refs. \cite{Song:2012cd,Xu:2012gf} the difference between particles and antiparticles
is explained by the presence of a vector
mean field potential which is repulsive
for particles and attractive for antiparticles. 
In \cite{Song:2012cd} it is introduced at the quark level within the Nambu-Jona-Lasinio
(NJL) model, while in \cite{Xu:2012gf}, at the hadronic level. 
As a consequence of these potentials, antiparticles
are attracted by the matter and
are trapped in the system, whereas particles feel
a repulsive force and have the tendency to leave the system
along the participant plane. With the potentials included, a fair qualitative
agreement was achieved \cite{Xu:2012gf}.

In Ref. \cite{Steinheimer:2012bn}, a hybrid (hydrodynamical plus Ultra-relativistic Quantum
Molecular Dynamics \cite{Bass:1998ca})
calculation was performed. 
The effect for the protons-antiprotons primarily results from 
the annihilation process.
Antiprotons moving in the out-of-plane  direction
encounter more protons to annihilate with than those
moving in the in-plane  direction.
Another effect discussed in this paper
is related to the event plane calculation. It was
claimed that fluctuations in this calculation can bias the
event plane to be rotated towards the most abundantly
produced particles. This would, for example, increase the
$v_2$ values for protons and reduce them for anti-protons.

The 3FD model proposes a more plausible (as compared to Refs. 
\cite{Song:2012cd,Xu:2012gf}) explanation of the difference
between particle and antiparticle elliptic flow in terms of three interacting fluids: 
the fireball, consisting of particles newly produced near the spacial center of the colliding system, 
and two baryon-rich fluid ($p$ and $t$) associated with   
leading particles which traversed the whole system and are finally located in 
longitudinally peripheral regions.
The particle-antiparticle difference explained by the increase of
nuclear stopping in heavy ion collisions with decreasing
energy. When the nuclear stopping becomes strong, the mid-rapidity quantities are determined  
not only by particles newly produced near the spacial center (the f-fluid)   
but also contributed by leading particles (the p- and t-fluids). 
The center and peripheral regions differently 
contribute to the mid-rapidity elliptic flow of different species, because 
they have different content of particles-antiparticles (quarks-antiquarks). 
This naturally results in different $v_2$ of particles and antiparticles
because the center and peripheral regions have different $v_2$ patterns. 
This explanation is similar in some features to those proposed in 
Refs. \cite{Dunlop:2011cf,Greco:2012hh}. Contrary to Refs. \cite{Dunlop:2011cf,Greco:2012hh}, 
the 3FD model calculates contribution from peripheral regions, 
rather than does assumptions on them, and does not employs the quark coalescence.

\section{Comparison with Data}
\label{Comparison}

The elliptic flow 
is proportional to the spatial anisotropy \cite{Voloshin:2008dg,Ollitrault92}. 
Usually, for this purpose one uses the eccentricity $\varepsilon$ 
defined by 
\begin{eqnarray}
 \label{eps}
\varepsilon = \frac{\langle y^2\rangle - \langle x^2\rangle}%
{\langle y^2\rangle + \langle x^2\rangle}\,\,.
\end{eqnarray}
Mean values of spacial transverse coordinates $\langle x^2\rangle$ 
(out of the reaction plane) and $\langle y^2\rangle$ (in the reaction plane)
are usually calculated
with either the wounded-nucleon~(WN) or the binary-collision~(BC) weights,
for details see Ref. \cite{Jacobs00}. 
These calculations are
based on the usual Woods--Saxon profile of nuclear density  
\begin{eqnarray}
 \label{Woods-Saxon}
\rho (r) = \frac{\rho_0}{1+\exp[(r-R_A)/d]}, 
\end{eqnarray}
where $\rho_0$ is the normal nuclear density, $R_A=1.12 A^{1/3}$ is the raduis of a nucleus with mass number $A$, 
and $d$ is a diffuseness of the nuclear surface.

As long as the eccentricity is small, elliptic flow should be directly proportional
to the eccentricity. For numerically
large eccentricities the direct proportionality could break in principle, but as was
shown in the very first hydrodynamic calculation by Ollitrault \cite{Ollitrault92} 
the proportionality holds well even for rather large values of $\varepsilon$.

\begin{figure}[thb]
\includegraphics[width=6cm]{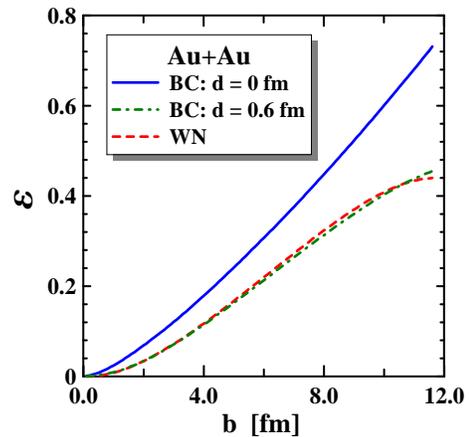}
\caption{
Spatial eccentricity
$\varepsilon$ as a function of impact parameter in Au+Au collisions
for different surface diffusenesses of the Au nucleus ($d$) and
different weights of averaging: the wounded-nucleon (WN) and
the binary-collision (BC) weights~\cite{Jacobs00}.
}
\label{fig0}
\end{figure}

Within the 3FD model the initial nuclei are represented by sharp-edged
spheres, i.e. the zero diffuseness ($d=0$). This is done for stability of 
the incident nuclei before collision. This circumstance essentially affects 
the eccentricity. 
The  results obtained with $d=0$ and the realistic value of $d=$ 0.6 fm 
calculated with BC  weights
are shown in Fig.~\ref{fig0}. As seen, the $(d=0)$-result noticeably exceeds 
the eccentricity for the physically realistic value of $d=$ 0.6 fm. 
Moreover, the ($d=0.6$ fm)-result with BN weights practically coincides with 
the eccentricity calculated with WN weights. The latter is 
considered as a realistic eccentricity and is accepted in the 
experimental analysis Ref. \cite{Jacobs00}.

\begin{figure*}[phtb]
\includegraphics[width=12.0cm]{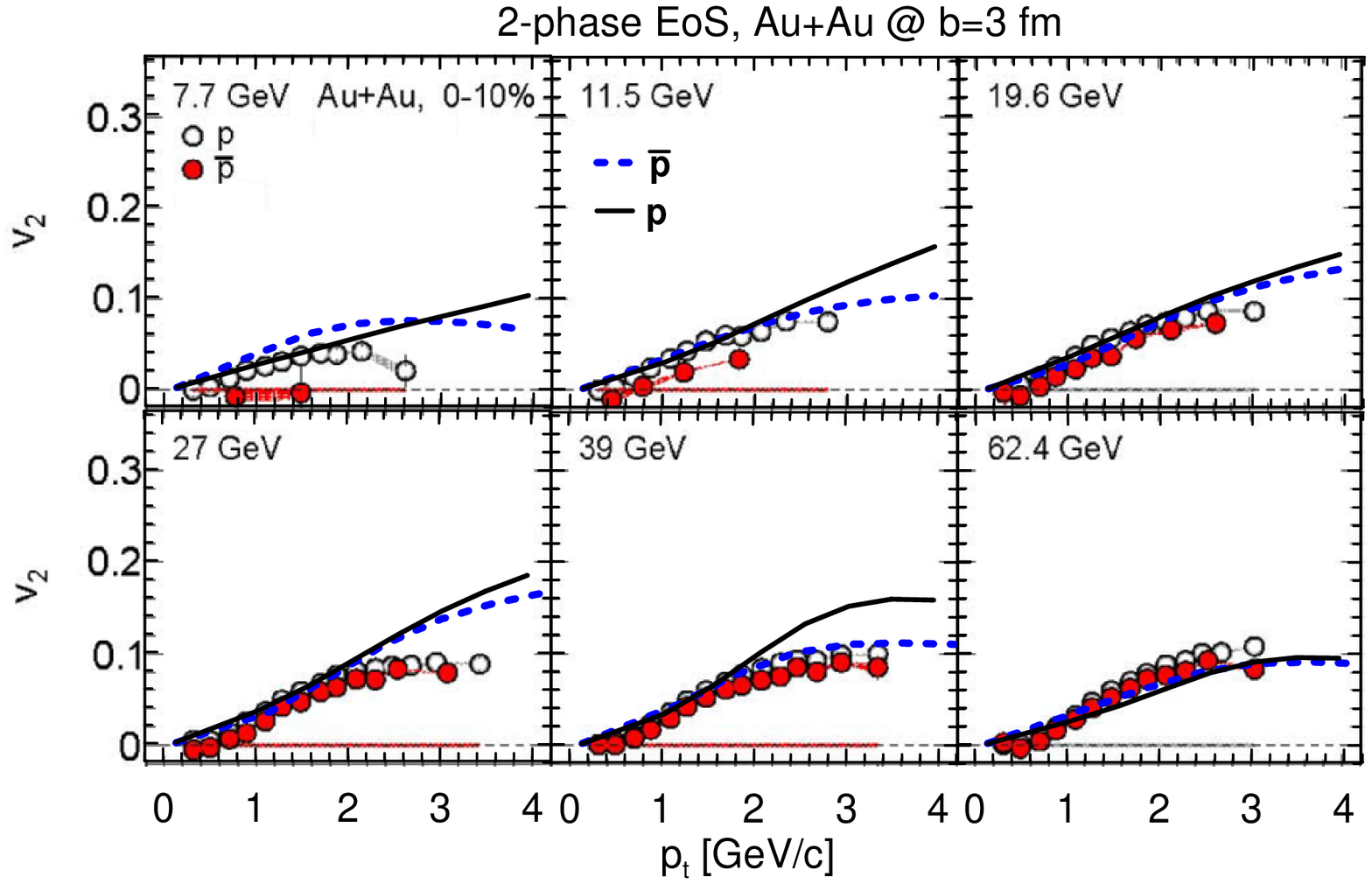}\\
\includegraphics[width=12.0cm]{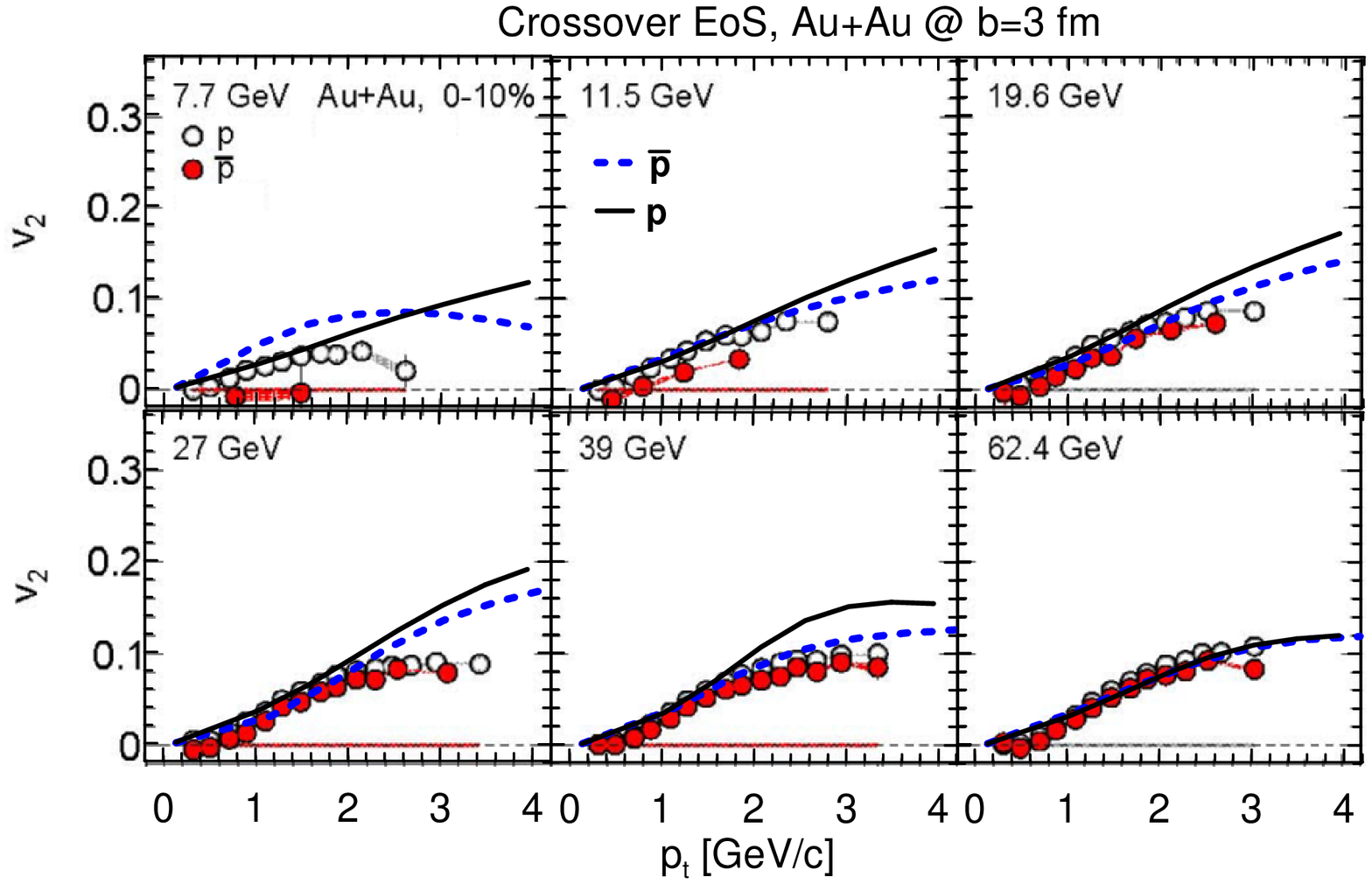}\\
\includegraphics[width=12.0cm]{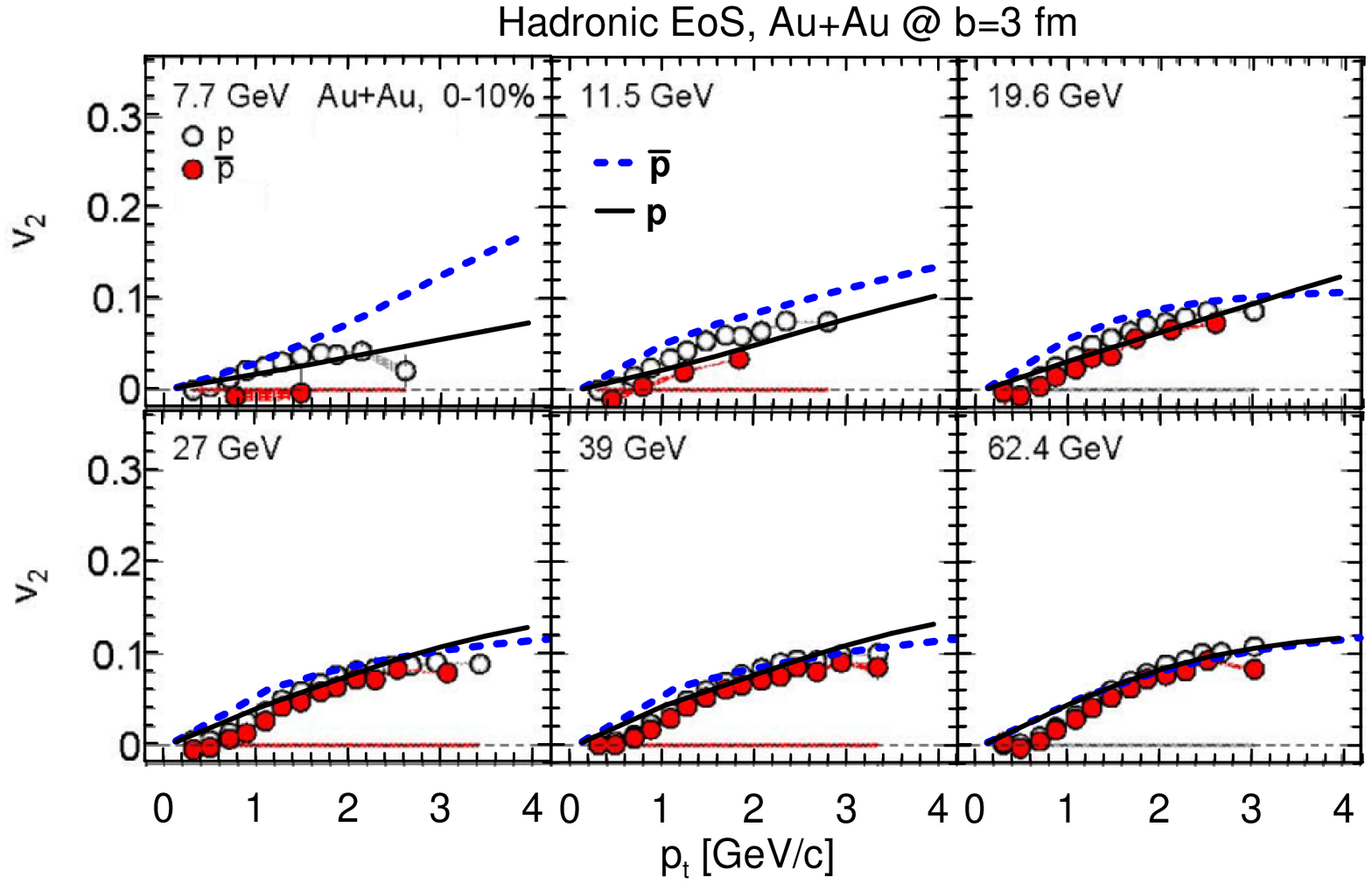}\\
 \caption{
Elliptic flow of protons and antiprotons   
at mid-rapidity in
central (0\%-10\%) Au+Au collisions at $\sqrt{s_{NN}}=$ 7.7, 11.5, 19.6, 27, 39 and 62.4  GeV as 
a function of transverse momentum. The 3FD calculations
are performed at $b=$ 3 fm.
Experimental data are from Star Collaboration \cite{Adamczyk:2013gw}. 
} 
\label{fig1}
\end{figure*}
\begin{figure*}[phtb]
\includegraphics[width=12.0cm]{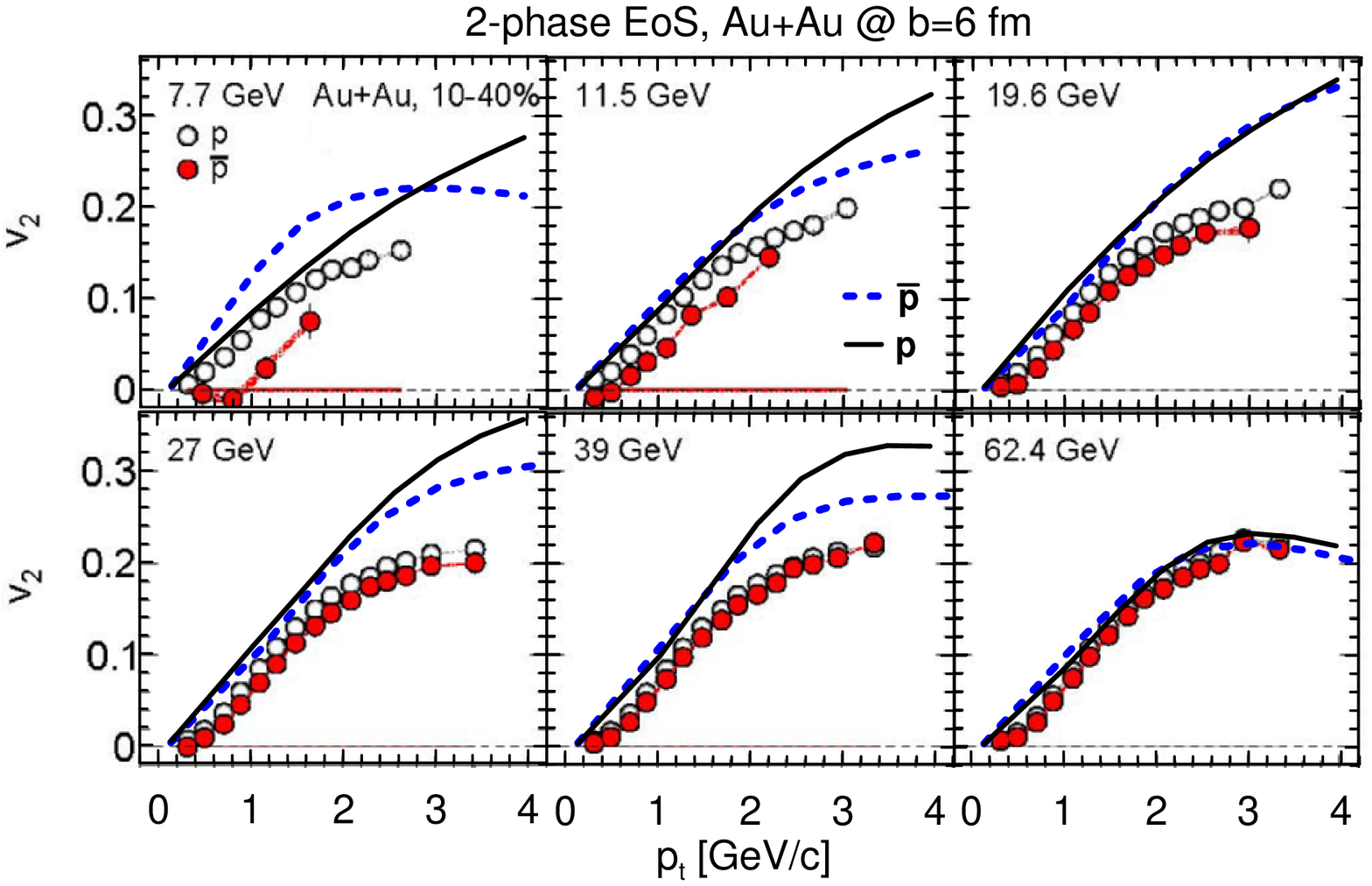}\\
\includegraphics[width=12.0cm]{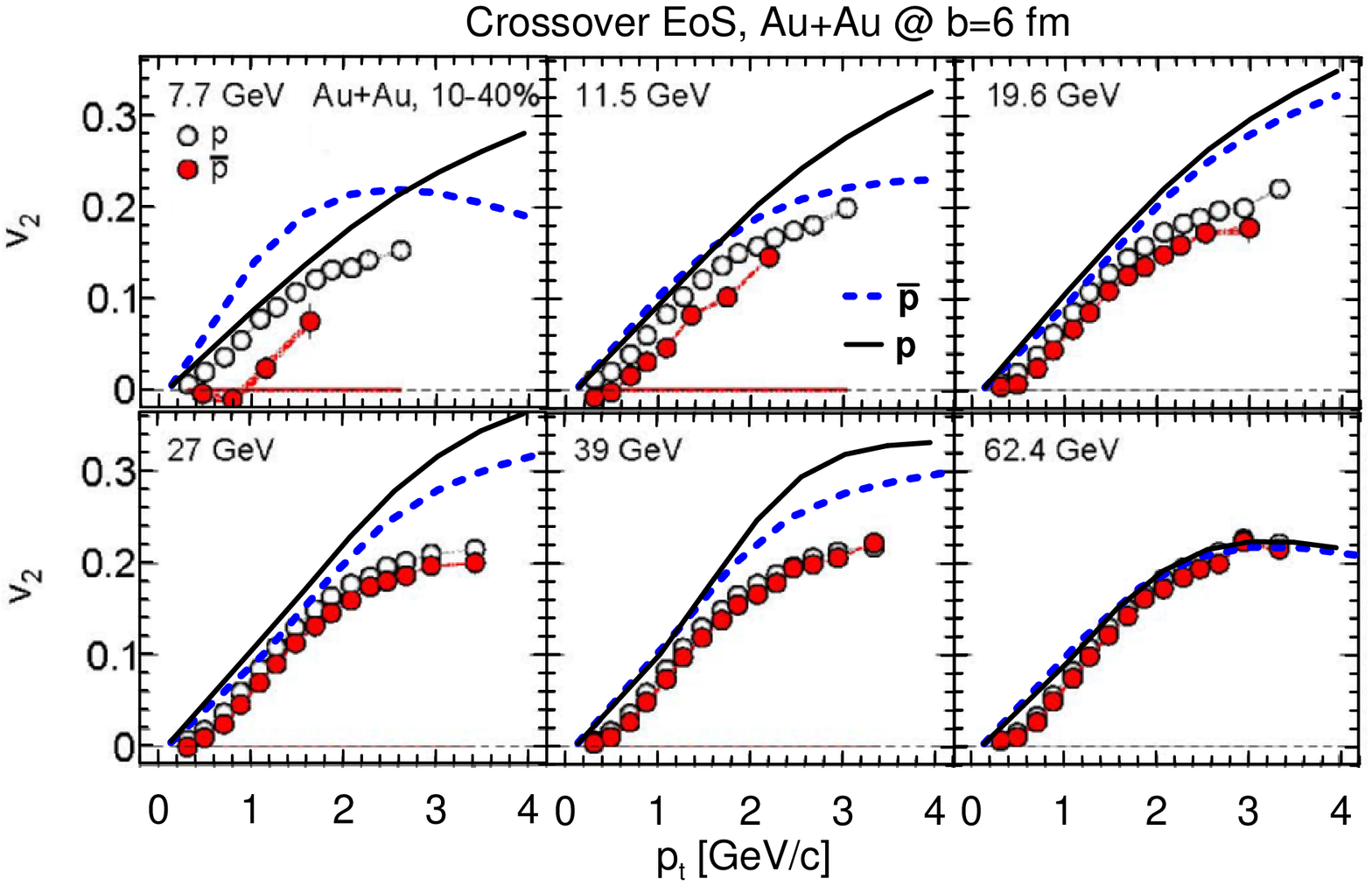}\\
\includegraphics[width=12.0cm]{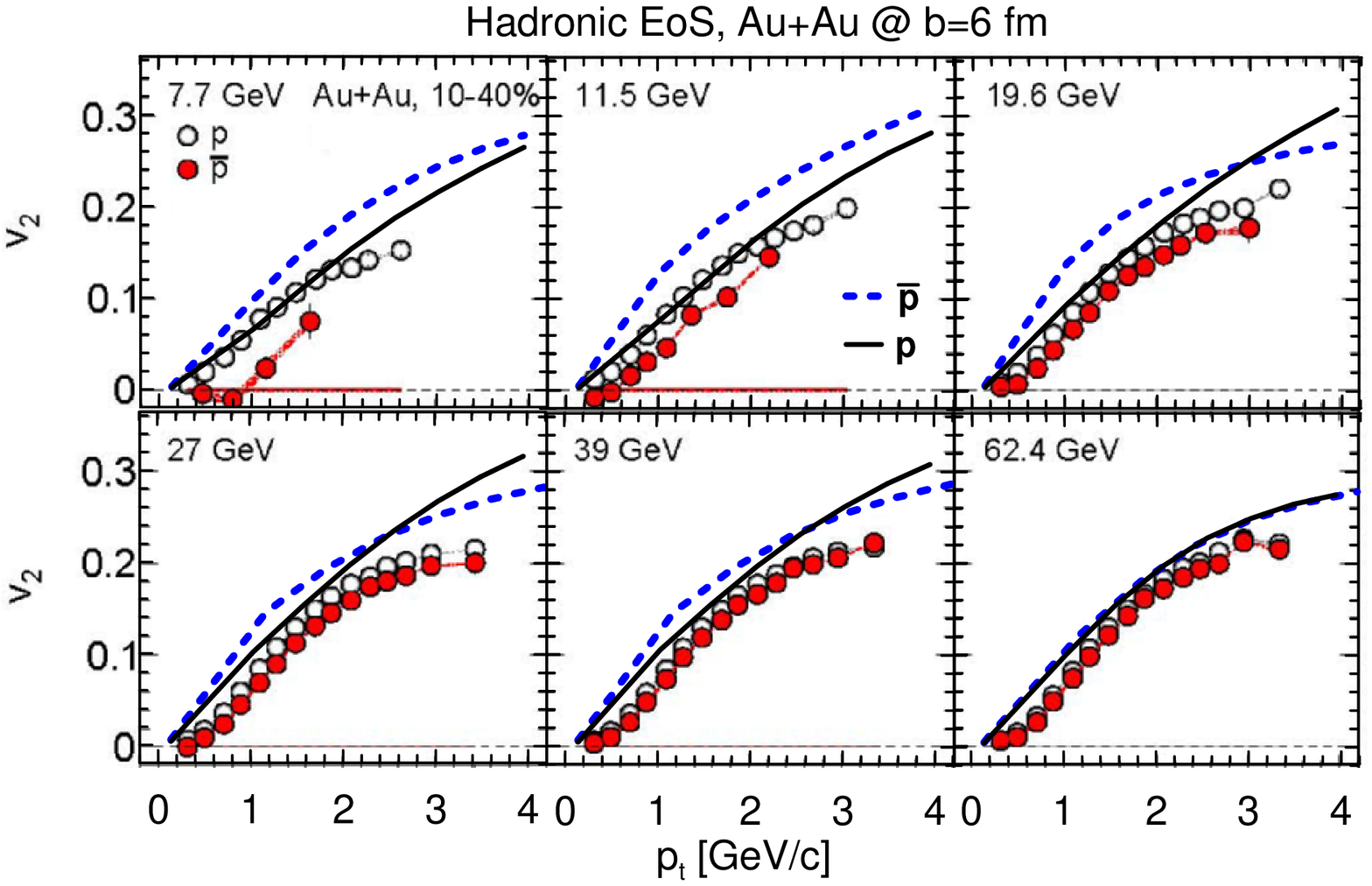}\\
 \caption{
The same as in Fig. \ref{fig1} but for 
mid-central (10\%-40\%) Au+Au collisions. The 3FD calculations
are performed at $b=$ 6 fm.
} 
\label{fig2}
\end{figure*}
\begin{figure*}[phtb]
\includegraphics[width=12.0cm]{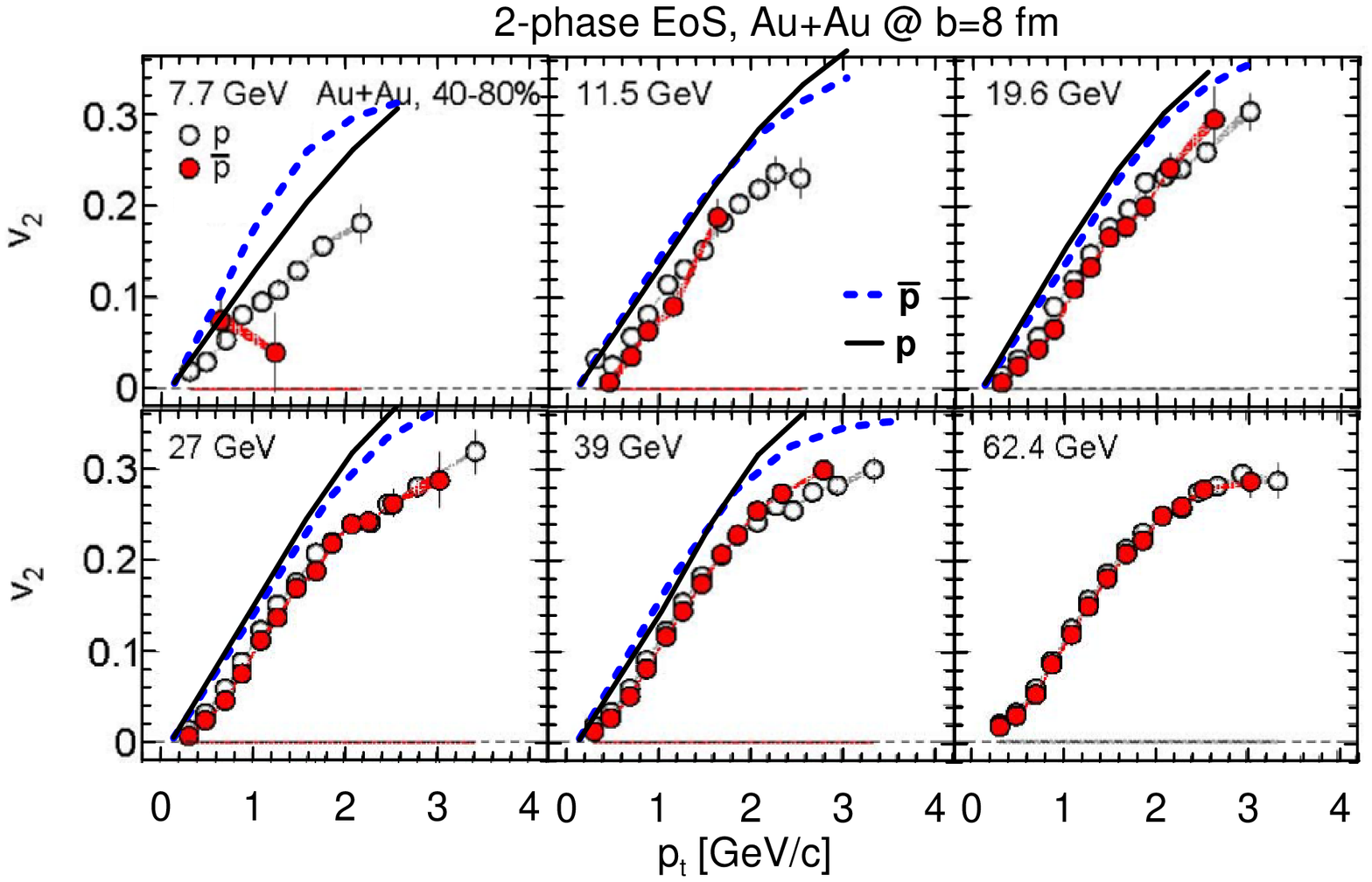}\\
\includegraphics[width=12.0cm]{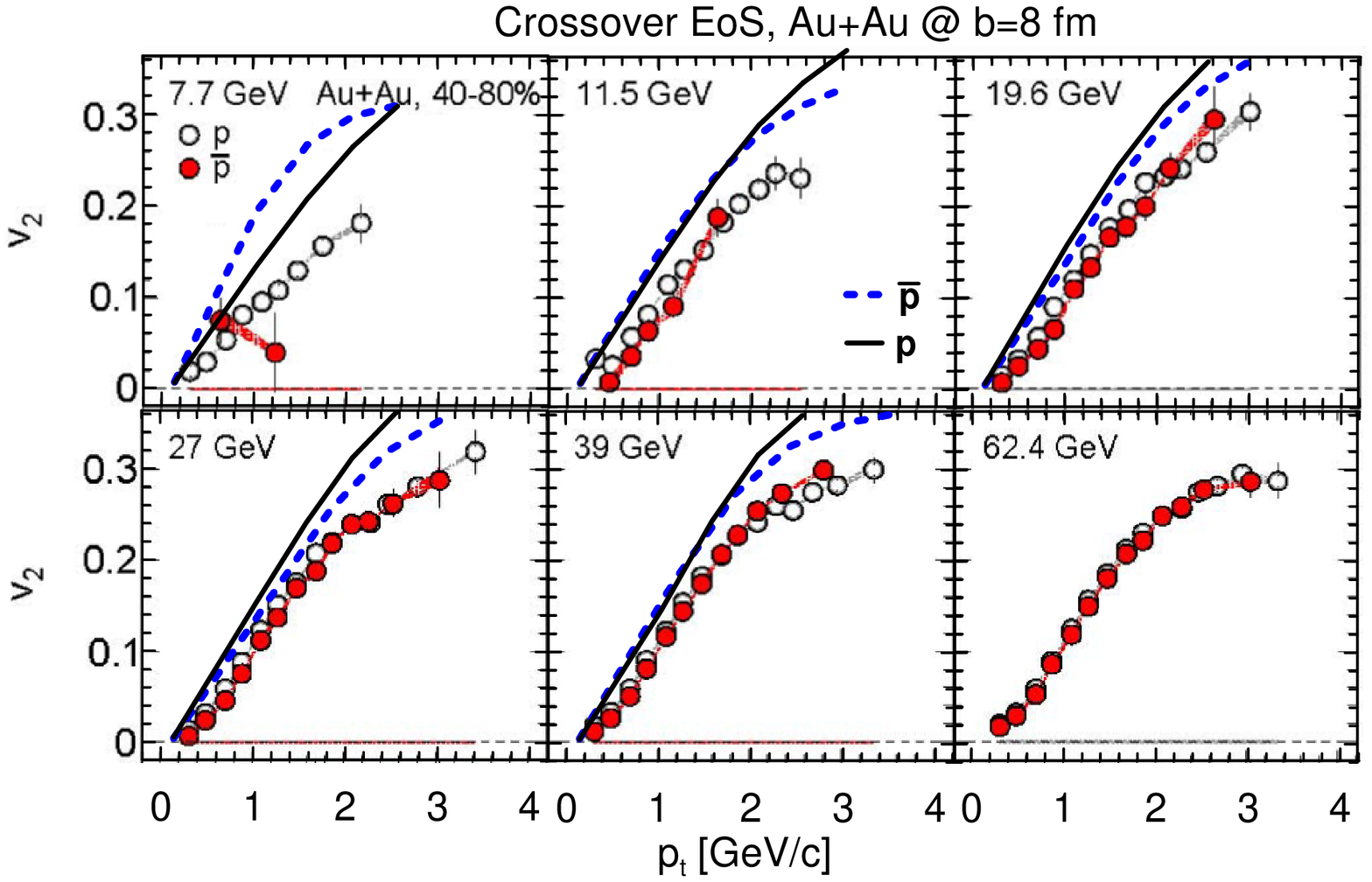}\\
\includegraphics[width=12.0cm]{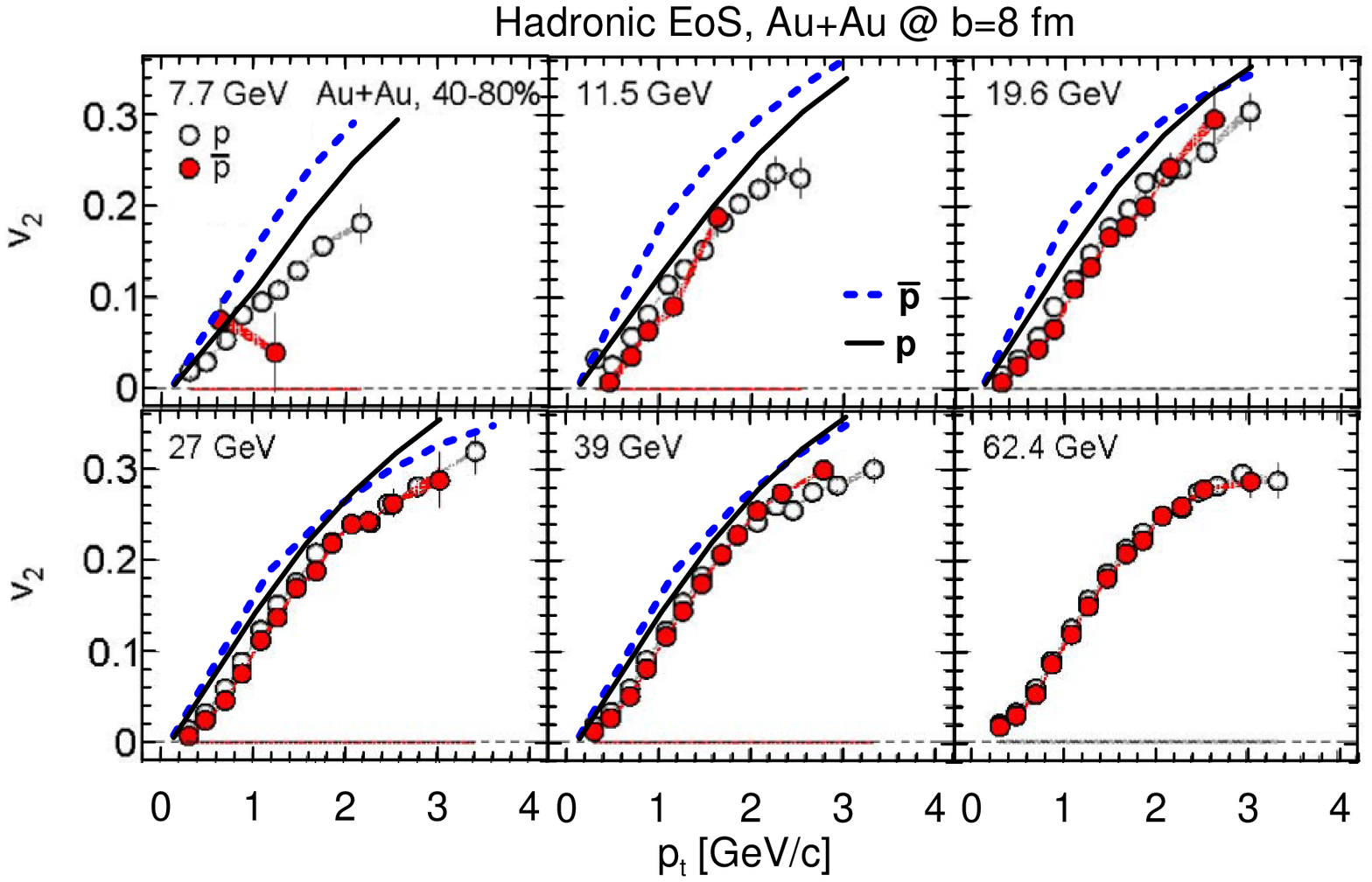}\\
 \caption{
The same as in Fig. \ref{fig1} but for peripheral
(40\%-80\%) Au+Au collisions. The 3FD calculations
are performed at $b=$ 8 fm.
} 
\label{fig3}
\end{figure*}

The overestimation of $\varepsilon$ in the 3FD model naturally results 
in the respective overestimation of the elliptic flow. There are two ways 
to compensate this overestimation: either the impact parameter of the collision 
should be reduced to get a reasonable value of $\varepsilon$ for a 
considered centrality bin or the calculated value of $v_2$ should be rescaled 
with the factor of $\varepsilon_{BN}(d=0.6 \mbox{ fm})/\varepsilon_{BN}(d=0)$, 
because $v_2\propto \varepsilon$, as mentioned above. The latter method 
is applied in this paper, i.e. all the $v_2$ values displayed below are 
rescaled with the above factor. 
As the calculations are performed at fixed impact parameters, 
only  proton and antiproton elliptic flow is considered for which 
data with comparatively narrow centrality selection are available \cite{Adamczyk:2013gw}.

Results of calculations of proton and antiproton elliptic flow at mid-rapidity 
and their comparison with STAR data \cite{Adamczyk:2013gw} for three centrality bins 
are presented in Figs. \ref{fig1}, \ref{fig2} and \ref{fig3}. 
The quality of data reproduction is quite reasonable for calculations at fixed impact parameters that 
are representative for the centrality bins considered. 
It should be kept in mind that in spite of the best reproduction at the top considered energy of 
62.4 GeV, results at this energy   
are still not quite accurate, since an accurate computation requires 
unreasonably high memory and CPU time.  Precisely due to this reason 
the calculations of peripheral ($b=$ 8 fm) collisions at 62.4 GeV have not been done.

All scenarios, including the purely hadronic one, quantitatively reproduce the proton data  
approximately to the same extent. It simply means that all scenarios represent strongly 
collective (fluid-like) behavior of the system that results in a large elliptic flow. 
The value of $v_2$ does not directly indicate either hadronic or partonic content 
of the matter. To distinguish between the hadronic and partonic content we need 
more delicate properties of the elliptic flow.
The calculated results 
as a rule overestimate the data. This makes room for the viscosity and/or afterburner corrections 
to the calculated results. In the 3FD model neither viscosity nor kinetic afterburner 
at the final stage of the expansion were incorporated.

Elliptic flows of protons and antiprotons at mid-rapidity coincide only if these 
are formed by the single fireball fluid, i.e. if baryon-rich fluids (the projectile and 
target ones) are well separated from the fireball fluid in rapidity. 
As seen this is almost the case at 62.4 GeV. 
With the incident energy decrease this separation in rapidity space becomes smaller, 
and as a result the the proton and antiproton $v_2$ differ from each other more and more. 
This is a consequence of difference in proton and antiproton content and 
in the $v_2$ patters of the baryon-rich  and  fireball fluids. 
Of course, exceptions from this rule are possible if $v_2$ patters of different fluids
turn out to be very similar by some reason. For instance, this is the case  
for the energy of 19.6 GeV within deconfinement scenarios.   
Typically, $v_2$  of protons and antiprotons become closer to each other with the incident 
energy rise, similarly to that observed in the experiment. This takes place for  
all considered scenarios because the baryon-rich  and  fireball fluids get more and more
separated in rapidity.

As for more delicate properties of the elliptic flow mentioned above, 
contrary to the hadronic scenario, all 
deconfinement scenarios result in proton elliptic flow exceeding or being approximately equal 
(with accuracy of the calculation) the antiproton one 
(with the exception of the lowest considered incident energy of 7.7 GeV). The latter is 
in qualitative agreement with the data. The lowest energy of 7.7 GeV is already 
quite close to onset of deconfinement within 2-phase-EoS and crossover-EoS scenarios 
\cite{Ivanov:2013wha,Ivanov:2012bh}. 
Therefore, the model predicts a kind of irregular behavior in this energy region 
that does not agree with the data  at present. 
Within deconfinement scenarios the antiproton $v_2$ data are, as a rule, better reproduced 
in the low-momentum region, that is the prime region of the applicability of the fluid 
description. This indicates the preference of the deconfinement scenarios.

\section{Conclusions}
\label{Conclusions}

The elliptic flow of produced particles is one of the most
sensitive observables that brings information about the
degree of collectivity during the expansion stage of heavy-ion collisions. 
When the collectivity is strong, like in
the case of ideal hydrodynamics, the elliptic flow takes the highest value
(the so called hydrodynamic limit). If the
collectivity is weak, like in a dilute system of weakly interacting particles, it is close
to zero.
The value of $v_2$ does not directly indicate either hadronic or partonic content 
of the matter. 
This is confirmed by approximately the same extent of reproduction of 
the proton and antiproton elliptic flow achieved within all scenarios, 
including the purely hadronic one.

To distinguish between the hadronic and partonic content we need 
more delicate properties of the elliptic flow. 
An indication in favor of the deconfinement scenarios is 
better reproduction of the antiproton elliptic flow
in the low-momentum region within these scenarios because  
this region is the main domain of the applicability of the fluid 
description.

In the present version of the 3FD model neither viscosity nor kinetic afterburner 
at the final stage of the expansion are incorporated. 
Therefore, it is not surprising that the calculated elliptic flow, 
as a rule, overestimates the data. This makes room for the viscosity and/or afterburner corrections 
to the calculated results.

The calculated elliptic flows of protons and antiprotons become closer to each other with the incident 
energy rise, similarly to that observed in the experiment. 
In fact, the elliptic flows of protons and antiprotons at mid-rapidity coincide only if these 
are formed by a single fluid, expanding in the center of the colliding system,  
in other words, if this ``center'' fluid is well separated in rapidity from 
projectile- and target-like leading particles, dominantly occupying peripheral rapidity regions. 
If the nuclear stopping rises, as it takes place with incident energy decrease, 
the rapidity gap between the ``center'' fireball and the leading-particles matter shrinks. 
Then the leading-particles matter starts to contribute to the mid-rapidity quantities. 
In its turn, this results in a difference between elliptic flows of protons and antiprotons
that is a consequence of difference in proton and antiproton content and 
in the $v_2$ patters of the ``center'' fireball and the leading-particles matter. 
In the 3FD model this mechanism is realized in terms of three interacting fluids: 
the fireball one, consisting of particles newly produced near the spacial center of the colliding system, 
and two baryon-rich fluid ($p$ and $t$) associated with   
leading particles which traversed the whole system and are finally located in 
longitudinally marginal regions.

\vspace*{3mm} {\bf Acknowledgements} \vspace*{2mm}

I am grateful to A.S. Khvorostukhin, V.V. Skokov,  and V.D. Toneev for providing 
me with the tabulated 2-phase and crossover EoS's. 
The calculations were performed at the computer cluster of GSI (Darmstadt). 
This work was supported by The Foundation for Internet Development (Moscow)
and also partially supported  by  
the Russian Ministry of Science and Education 
grant NS-215.2012.2.


\end{document}